# Resurrection of the Melting Line in the Bose Glass Superconductor

Leo Radzihovsky

*The James Franck Institute and Physics Department, University of Chicago, Chicago, IL 60637*

(March 26, 1995)

I argue that in contrast to $B < B_\phi$, where the Vortex Liquid (VL) freezes into the Strongly-pinned Bose Glass phase (SBG), with flux-line vortices localized by the columnar defects, for $B > B_\phi$, the additional vortices see a significantly weaker random $z$-independent potential due to the caging by the vortices strongly pinned on the columnar defects. The result is that for $B > B_\phi$, upon cooling, the VL undergoes a sharp crossover to the Interstitial Liquid (IL) where SBG coexists with a resistive liquid of interstitial vortices. At lower $T$ the IL freezes into a distinct Weakly-pinned Bose Glass phase (WBG). Therefore for $B > B_\phi$ the true linear superconductor only forms at $T$ well below the $B < B_\phi$ irreversability line.

In high-$T_c$ superconductors, because of the enhanced fluctuations, as compared to conventional superconductors, over a large portion of the B-T phase diagram, the Abrikosov lattice melts [1], [2] into a resistive Vortex Liquid (VL) of highly mobile flux lines. At lower temperature and magnetic field, below the irreversibility line, depending on the nature of the disorder, the VL freezes into one of the proposed new glassy phases, leading to true linear superconductivity. In the presence of point disorder such as oxygen vacancies and interstitials, Vortex Glass (VG) [3] has been proposed as the low-temperature phase replacing the Abrikosov lattice.

With the success of Civale, et al. [4] in significantly shifting the irreversibility line to higher $T$ and $B$ by irradiating the superconductor with heavy energetic ions, much of recent attention has focussed on flux-line vortices in the presence of correlated disorder such as columnar defects. It was previously known that the interacting flux lines map onto the quantum mechanics of two-dimensional bosons [1]. Key progress was made when Nelson and Vinokur (NV) realized that the random columnar defects naturally map onto a time-independent random potential for the bosons [5], resulting in the previously studied problem of two-dimensional quantum bosons on a random substrate [6]. NV interpreted the irreversibility line as the phase transition where the VL freezes into a superconducting Bose Glass (BG) of vortices localized on columnar defects. The mapping led to a detailed analysis of the BG transition and of the properties of the low temperature BG phase.

Most of the analysis of Ref. [5] had focussed on the regime $B < B_\phi$ with the matching field $B_\phi = \phi_o/d^2$ corresponding to an average spacing $d$ between the columnar defects. In their work, however, NV did point out, that since the random potential for $B > B_\phi$ is also $z$-independent, the BG will exist in this regime as well. In this Letter, I will reexamine in more detail this localization problem, contrasting the regimes $B < B_\phi$ and $B > B_\phi$. I will argue for a quantitatively modified phase diagram of the BG superconductor for $B > B_\phi$, with a potentially sharp crossover as a function of $T$ or a 1st-order transition, from the VL phase to the Interstitial Liquid (IL), and a strong suppression of the BG transition (the irreversibility line) to lower $T$, down to where the melting line (in the absence of columnar defects) used to be (see Fig.1). Although I do not offer many new technical details (most of these can be found in Refs [6,5]), my arguments have concrete physical consequences that are experimentally verifiable. The important consequence of my work is that the introduction of columnar defects should be much less effective in increasing the pinning, raising the critical currents, and shifting the irreversibility line to higher $B$ and $T$ in the $B > B_\phi$ regime, in contrast to the $B < B_\phi$ regime.

The simplest standard model that captures the physics of $N$ interacting flux-line vortices through a sample of thickness $L$, in the presence of columnar defects, is described by the free energy $F$,

$$F[\vec{r}_i(z)] = \frac{\tilde{\epsilon}_l}{2} \sum_{i=1}^{N} \int_0^L dz (d\vec{r}_i/dz)^2 + \qquad (1)$$

$$+ \sum_{i>j=1}^{N} \int_0^L dz V[\vec{r}_i(z) - \vec{r}_j(z)] + \sum_{i=1}^{N} \int_0^L dz U[\vec{r}_i(z)],$$

where $\vec{r}_i(z)$ is the local deviation of the $i$th vortex from the applied field $z$-direction, $V(r) = 2\epsilon_0 K_0(r/\lambda_{ab})$ is the interaction, which is logarithmic up to the London penetration length $\lambda_{ab}$, and falls off exponentially for $r > \lambda_{ab}$. $U(r)$ is the random potential of range $b \approx 70\text{\AA}$, induced by the columnar defects, with disorder-averaged variance $\overline{U(\vec{r})U(\vec{r'})} = 2U_0^2(b^4/d^2)\delta^{(2)}(\vec{r} - \vec{r'})$. The line tension $\tilde{\epsilon}_l = (M_{ab}/M_z)\epsilon_0 \ln(\lambda_{ab}/\xi)$ is significantly reduced relative to the condensation energy $\epsilon_0 = (\phi_0/4\pi\lambda_{ab})^2$ due to the considerable anisotropy of high-$T_c$ materials ($M_{ab}/M_z \approx 1/100$).

As discussed in detail by NV [5], building on the ideas of Ref. [6], the above model for $B < B_\phi$ predicts three distinct vortex phases. I first remind the reader of the properties of these phases and then turn to the new predictions, relying on many results of NV.

At high temperature and magnetic field, the fluctuating vortices can move freely through a sample, "diffusing" along the $z$-axis, $\overline{<|\vec{r}_i(z)-\vec{r}_i(z)|^2>} \approx 2zT/\tilde{\epsilon}_l$. This leads to a resistive VL phase (corresponding to the superfluid phase of the quantum bosons), with the flux-flow resistivity described by the Bardeen-Stephen (BS) formula, $\rho_{ff} = \rho_n B/B_{c2}$; $\rho_n$ is the normal-state resistivity. The bulk modulus $c_{11}$ and the tilt modulus $c_{44} \sim \tilde{\epsilon}_l B$ are finite, while the shear modulus vanishes in this phase. For $B < B_\phi$, as $T$ or $B$ is lowered below $T_{BG}(B)$, the BG transition takes place, at which the flux lines localize on the columnar defects, as described by the *finite* transverse wandering length $l_\perp$ and $\overline{<|\vec{r}_i(z)-\vec{r}_i(z)|^2>} = l_\perp^2$. Upon approaching the transition the tilt modulus (proportional to the inverse superfluid density for the quantum bosons) will diverge leading to the transverse Meissner effect, requiring a finite tipping angle $\theta_c \approx (U_0 b^2 n/d^2 c_{44})^{1/2}$ of the external field (relative to the columnar defects) to change the average internal magnetic field direction.

At the BG transition, postulating two divergent length scales, the localization length $l_\perp \sim 1/(T_{BG}-T)^{\nu'}$, the correlation length along the $z$-direction $l_\parallel \sim l_\perp^2$, and the corresponding divergent relaxation time $\tau \sim l_\perp^{z'}$, one can construct the BG scaling theory. This analysis predicts the I-V curve given by $E(J) = l_\perp^{-1-z'} F_{+/-}(Jl_\perp l_\parallel \phi_0/cT)$, with the scaling function $F_{+/-}(x)$, which in the VL phase ($T > T_{BG}$) is $F_+(x) \sim x$, leading to a finite linear resistivity $\rho_{ff} \sim (T - T_{BG})^{\nu'(z'-2)}$. In the truly superconducting BG phase, $\rho_{ff}$ vanishes and various transport regimes lead to highly nonlinear I-V characteristics. At small $J$, in the variable-range-hopping regime, where the vortices hop to distant columnar defects via superkinks, $E(J) \sim \exp(-1/J^{1/3})$. At high $J$, closer to the critical current $J_c$, $E(J) \sim \exp(-1/J)$, and at the BG transition $E(J) \sim J^{(1+z')/3}$. The scaling theory also predicts the existence of $\theta_c \sim (T_{BG}-T)^{3\nu'}$, which leads to a cusp in resistivity $R(\theta)$ and in the $T_{BG}(H_\perp)$, as is observed in experiments on crystals with correlated pinners [9,10], the features that distinguish BG from the isotropic VG phase. Finally, as pointed out by NV, a highly correlated Mott Insulator (MI) phase might appear as $B \to B_\phi$, in which both $c_{44}$ and $c_{11}$ are infinite.

I now closely examine the situation for $B \geq B_\phi$ (see Fig.1). Because of the divergence of the bulk modulus $c_{11}$, the MI phase is expected to exhibit rigidity to the introduction of any additional flux lines, and in this way behaves similarly to the Meissner phase. In fact, recent magnetization-relaxation experiments appear to verify its existence, observing a drastic suppression of the magnetization relaxation rate for $B \approx B_\phi$ [8].

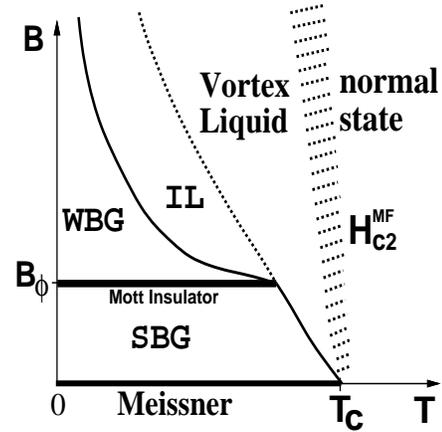

FIG. 1. The proposed phase diagram for the BG superconductor. As discussed in the text, the SBG phase, where the vortices are strongly pinned by columnar defects, is replaced by two phases: the resistive IL and the superconducting WBG phase for $B > B_\phi$. The superconducting MI phase is the Meissner phase for the interstitial vortices.

As the applied field is increased above $B_\phi$, additional vortices begin to enter the sample. It is energetically more favorable for these vortices to occupy the interstitials between the vortices already strongly pinned by the columnar defects. At very low temperatures these interstitial vortices will also be localized by the repulsion $V(r)$ from the vortices on columnar defects, caged-in by the nearest neighbors. Since the randomly localized columnar vortices will appear to the interstitial flux lines as a random, $z$-independent potential, I expect that the interstitial vortices will also freeze into the BG phase [5].

The resulting Weakly-pinned Bose Glass (WBG) phase will have a vanishing linear resistivity, and the transport describable by the BG scaling theory. In fact, the equilibrium properties of this phase should be similar to that of the SBG, described above, in all respects except for the pinning energy scale which determines $T_{BG}(B)$ and $J_c$. In the SBG phase (at low $T << T_{SBG}^* \approx 70$ K, where the effects of thermal renormalization are not important), the columnar defect pinning energy is determined by $U_0 \approx \epsilon_0$ with the corresponding $J_c^{SBG} = cU_0/\phi_0 b$. Quantitatively I expect that the interstitial vortices controlling the formation of the WBG phase are significantly weaker localized. Because the repulsive energy $V(r)$ and the columnar defect localization energy $U(r)$ are both of the order of the condensation energy $\epsilon_0$, it is difficult to have a good model-independent estimate of the localization energy scale of the interstitial vortices. One possible guess is $U_i \sim \tilde{\epsilon}_l/dn^{1/2} << U_0$, which leads to a reduced $J_c^{WBG} = cU_i/\phi_0 d << J_c^{SBG}$. Physically it is clear that $U_i << U_0$ because of the competition between the repulsion from the neighboring vortices strongly localized on columnar defects and the "attraction" by the columnar defects. One limit in which it is unambiguously clear that $U_i << U_0$ is when the spacing between the columnar de-fects $d >> \lambda_{ab}$, in which case $U_i \approx U_o e^{-d/\lambda_{ab}} << U_o$. In fact, recent experiments of Ref. [8], which infer from the size of the hysteris loop two distinct critical currents for $B < B_\phi$ and $B > B_\phi$ with $J_c^{SBG} \approx 2J_c^{WBG}$, give strong experimental support to my arguments. The existence of two critical currents will also lead to a double-sloped spatial magnetic field profile of the Bean critical state [7].

Given this separation of the pinning energy scales for $B > B_\phi$, in exploring the physics beyond linear response it is convenient to treat the vortex system in a two-fluid model spirit, with the WBG phase consisting of vortices strongly pinned on the columnar defects and a fraction of interstitial vortices localized by the repulsive interactions. This two-fluid model then leads to the shape of the I-V curve which is significantly different from the SBG phase for $B < B_\phi$. As illustrated in Fig.2, the I-V curve is characterized by two depinning transitions at $J_c^{WBG}$ and $J_c^{SBG}$ to the flux-flow regime approaching that of the IL and VL, respectively.

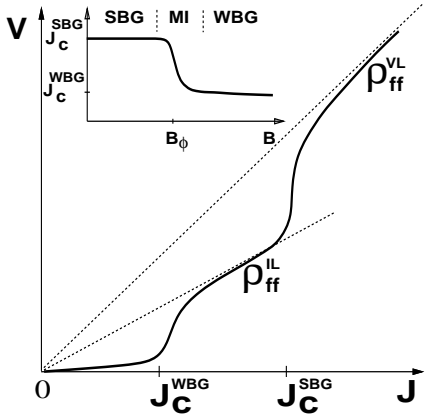

FIG. 2. The expected I-V curve in WBG phase, illustrating two depinning transitions, characterized by $J_c^{WBG}$ and $J_c^{SBG}$ and corresponding asymptotic flux flow resistivities. The inset shows qualitatively $J_c(B)$, observed in Ref.8.

The above arguments imply that for $B > B_\phi$, upon increasing temperature, the weakly-pinned interstitial vortices will be first to delocalize. When their transverse wandering exceeds their separation the WBG phase will melt (most likely via the BG transition) into the resistive IL. The resulting IL phase will have properties intermediate between that of the SBG and the VL, calculable within the simple two-fluid model. In this regime the interstitial vortex fraction is completely delocalized, while the $B_\phi/B$ fraction of flux lines remains pinned along the columnar pinners for a significant longer distance along the defect. This distinction between the strongly pinned columnar vortices and the interstitial liquid vortices is blurred by the delocalization of columnar vortices at long lengths and time scales and by exchange processes between the vortices. Therefore from the point of view of the symmetry of the order parameter the IL is identical to the VL phase and two are just different regimes of the same normal metallic phase. Another real possibility is that a 1st order transition separates IL phase from the VL phase, which would lead to a jump in quantities (e.g. linear resistivity) that depend on the density of mobile vortices. In either case, the IL will exhibit quantitatively observable differences in behavior from the usual VL. The linear flux-flow resistivity in the IL is determined primarily by the interstitial vortex fraction, and is given by the generalized Bardeen-Stephen formula

$$\rho_{ff}^{IL} = \rho_n(B - B_\phi)/B_{c2} \qquad (2)$$

It is possible that fluctuations will modify the dependence on $B - B_\phi$ to another power law, but $\rho_{ff}^{IL}$ will vanish as the MI phase is approached.

As I discussed above, in the IL a mobile fraction $(B - B_\phi)/B$ of vortices is in the interstitials and the remaining vortices $B_\phi/B$ are localized on defects, with a constant dynamic exchange between the two "species". Since the vortices are really identical, each flux line spends the corresponding fractions of time in the interstitials and localized on columnar defects. During the fraction of time that the flux line is localized on a defect no diffusion takes place. Therefore the transverse "diffusion" coefficient $D_z = T/\tilde{\epsilon}_l$, determining the transverse wandering length $l_\perp(z) = \sqrt{2D_z z}$, is significantly renormalized in the IL regime to $D_z^{IL} = D_z(B - B_\phi)/B$. Given the fact that the tilt modulus $\tilde{\epsilon}_l$ maps onto the inverse of the superfluid density in the corresponding problem of quantum bosons I find that in the IL the corresponding superfluid density appears to be reduced by the $(B - B_\phi)/B$ fraction. The phenomenology of the IL is therefore quite similar to the superfluid Helium on a random substrate. There one observes that the low $T$ regime of the superfluid phase appears to have a substantially reduced superfluid density as compared to the extrapolation from the high $T$ regime and the observations are well described in terms of a superfluid fraction of He[4] on top of a "dead layer" of nonsuperfluid He[4]. The mapping of the IL onto the 2d quantum bosons therefore maps the interstitial vortices onto the superfluid fraction and interprets the observed "dead layer" in terms of the vortices strongly pinned on columnar defects. I therefore expect that this phenomenon should be observable directly in the flux-line vortex system.

It is expected that Bose Glass can be melted by tipping the applied field relative to the columnar defect $z$-direction [5]. However, unlike the case for $B < B_\phi$, two critical tipping angles $\theta_c^{WBG} \propto \sqrt{U_i(B - B_\phi)} << \theta_c^{SBG} \propto \sqrt{U_0 B}$ will appear, corresponding to $T_{WBG}(H_\perp)$ and $T_{SBG}(H_\perp)$ phase boundaries. At the lower boundary the WBG melts into the IL, with the interstitial vortex fraction $\nu = (B - B_\phi)/B$ depinning first, followed by the complete depinning of the columnar vortices at $\theta_{SBG}$, crossing over (or making a 1st order transition) into the VL. In the IL phase I expect to see striking features in the resistivity as a function of tipping angle, $\rho(\theta)$, for $J$ perpendicular to $z$ (similar signatures will appear in the WBG phase). For $\theta < \theta_c^{SBG}$, $\rho(\theta)$ will follow the usual $\rho(\theta) \approx \nu \rho_{ff}^{VL} sin^2(\theta)$ behavior, which will abruptly switch to a higher curve $\rho(\theta) \approx \rho_{ff}^{VL} sin^2(\theta)$ for $\theta > \theta_c^{SBG}$, as illustrated in Fig.3.

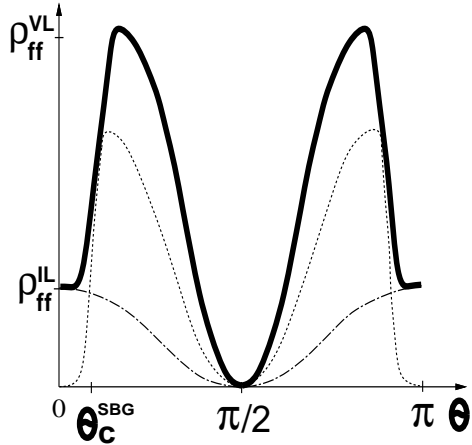

FIG. 3. The expected dependence of resistivity in IL phase, as a function of the tipping angle with $J \perp z$ (full curve). The dot-dashed curve is the smooth $sin^2(\theta)$ dependence of the IL fraction and the dotted curve is the response of the vortex fraction which is pinned by columnar defects up to the critical angle $\theta_c^{SBG}$.

Unlike the transition for $B < B_\phi$, for $B > B_\phi$, starting from the VL and lowering temperature, I predict a potentially sharp crossover or 1st order transition to the IL (at $T_{IL}$), followed by the BG transition to the WBG (at $T_{WBG}$). Given the two pinning energy scales, the transition temperatures can be roughly estimated as the temperature at which the thermally renormalized localization length $l_\perp(T) \approx l_0 (T/T^*)^2$ reaches the order of the corresponding vortex line separation $n^{-1/2}$, where $l_0$ is $d$ and $b$ for interstitial and columnar vortices, respectively. The thermal crossover temperatures in the two phases for $B > B_\phi$ are given by $T^*_{WBG} \approx (2\tilde{\epsilon}_l d^2 U_i)^{1/2}$ and $T^*_{SBG} \approx (2\tilde{\epsilon}_l b^2 U_0)^{1/2}$ leading to $T_{WBG}(B) \approx T^*_{WBG}(\phi_0/d^2(B - B_\phi))^{1/4}$ and $T_{IL}(B) \approx T^*_{SBG}(\phi_0/b^2 B)^{1/4}$. The transition from the VL to the IL will manifest itself as an additional plateau in the resistivity $\rho(T)$ which will disappear for $B < B_\phi$.

It is clear that my arguments strongly rely on the separation between the interaction $U_i$ and columnar $U_0$ pinning energy scales. I therefore present additional experimental evidence supporting the physical picture advocated above. Before the irradiation by heavy ions (i.e. in the virgin sample without columnar defects) the irreversibility line was the freezing/melting line into the Abrikosov lattice (possibly modified by weak point disorder at very long scales, which I ignore here). This freezing transition is determined by the vortex interaction energy scale which I expect to be similar to $U_i$. However, the experiments of Civale, et al. [4] find that the melting line is significantly shifted to higher temperature and magnetic field after the introduction of columnar defects, with the new energy scale determined by the columnar defect pinning energy $U_0$. I believe that this observed shift is the lower bound on the difference in pinning energy of the interstitial and the columnar vortices. In this sense, for $B > B_\phi$, the $T_{WBG}(B)$ transition from the IL to the WBG is the resurrection of the virgin melting line, smeared to a second-order transition by the columnar disorder.

In summary, I have presented arguments for the existence of the Weakly-pinned Bose Glass and the Interstitial Liquid phases for $B > B_\phi$, and described their phenomenology, which is quantitatively quite different from the Strongly-pinned Bose Glass and the conventional Vortex Liquid phases in the $B < B_\phi$ regime. I have discussed some of the experimentally observable features in the IV-curves and in tilting experiments that are unique to these new regimes. The observation of a shift of the irreversibility line to higher $T$ and $B$ for $B < B_\phi$ when columnar defects are introduced, and the lack of the corresponding shift for $B > B_\phi$ is a direct consequence of the presented picture.

I acknowledge helpful discussions with T. Rosenbaum, K. Beauchamp, M. P. A. Fisher, L. Balents and D. Nelson. This research was supported by the NSF (DMR $94 - 16926$), through the Science and Technology Center for Superconductivity.

[1] D. R. Nelson, Phys. Rev. Lett. **60**, 1973 (1988). D. R. Nelson and H. S. Seung, Phys. Rev. B **39**, 9153 (1989).
[2] H. Safar, et al., Phys. Rev. Lett. **69**, 824 (1992).
[3] M. P. A. Fisher, Phys. Rev. Lett **62**, 1415 (1989). D. S. Fisher, M. P. A. Fisher, and D. A. Huse, Phys. Rev. **B43**, 130 (1991).
[4] L. Civale, et al., Phys. Rev. Lett. **67**, 648 (1991).
[5] D. R. Nelson and V. M. Vinokur, Phys. Rev. Lett. **68**, 2398 (1992); Phys. Rev. B. **48**, 13060 (1993).
[6] M. P. A. Fisher, P. B. Weichman, G. Grinstein, and D. S. Fisher, Phys. Rev. B **40**, 546 (1989).
[7] K. M. Beauchamp, L. Radzihovsky, L. Shung, T. F. Rosenbaum, U. Welp, G. W. Crabtree, University of Chicago preprint.
[8] K. M. Beauchamp, et al., University of Chicago preprint.
[9] T. K. Worthington et al., Physica **153 C**, 32 (1988).
[10] G. W. Crabtree, et al., Proceedings of M$^2$-HTSC IV, Grenoble (1994).